\documentstyle[11pt,epsfig]{article}

\def\appendix{\par
 \setcounter{section}{0}
 \setcounter{subsection}{0}
 \def\thesection{Appendix \Alph{section}}
 \def\theequation{\Alph{section}.\arabic{equation}}
 \setcounter{equation}{0}}

\def\le{\left(}
\def\ri{\right)}

\def\no{\nonumber}

\def\rar{\rightarrow}

\def\e{\epsilon}
\def\f12{\frac{1}{2}}

\def\pd{\partial}

\begin{document}
\begin{titlepage}
\vskip 2cm
\begin{center}
{\Large \bf Fourier transforms of UD integrals}\\
\vskip 1cm  
Igor Kondrashuk$^{a}$ and Anatoly Kotikov$^{b}$\\
\vskip 5mm  
{\it  (a) Departamento de Ciencias B\'asicas, \\
Universidad del B\'\i o-B\'\i o, Campus Fernando May, Casilla 447, Chill\'an, Chile} \\
\vskip 1mm
{\it  (b) Bogoliubov Laboratory of Theoretical Physics, Joint Institute for Nuclear Research, \\
Dubna, Russia} \\
\end{center}
\vskip 20mm
\begin{abstract}
UD integrals published by N. Usyukina and A. Davydychev in 1992-1993 are integrals corresponding to ladder-type Feynman diagrams. 
The results are UD functions $\Phi^{(L)},$ where $L$ is the number of loops. They play an important role in ${\cal N}=4$ supersymmetic Yang-Mills theory. 
The integrals were defined and calculated in the momentum space. In this paper the position space representation of UD functions is investigated. 
We show that Fourier transforms of UD functions are UD functions of space-time intervals but this correspondence is indirect. 
For example, the Fourier transform of the second UD integral is the second UD integral.

\vskip 1cm
\noindent Keywords: UD integrals, UD functions.
\end{abstract}
\end{titlepage}

\section{Introduction}

As has been shown in Refs. \cite{Cvetic:2004kx} - \cite{Kondrashuk:2000qb}, Slavnov-Taylor identity predicts that the correlators 
of dressed mean fields for ${\cal N}= 4$ supersymmetric Yang-Mills theory in the position space can be represented in terms of UD integrals. 
The UD integrals correspond to the momentum representation of ladder 
diagrams and were calculated in Refs. \cite{Usyukina:1992jd,Usyukina:1993ch} in the momentum space, and the result can be written in terms of certain functions (UD functions) 
of conformally invariant ratios of momenta. Indeed, the $Lcc$ correlator in the position space in Wess-Zumino-Landau gauge of maximally supersymmetric 
Yang-Mills theory is a function of Davydychev integral $J(1,1,1)$ at two loop level 
\footnote{In the position space Feynman diagrams contain integrations over coordinates of internal vertices. 
Integration over internal vertices appears in dual representation of the momentum diagrams too (see below and \cite{Kazakov:1986mu,Kazakov:1987jk}).} 
\cite{Cvetic:2006iu,Cvetic:2007fp,Cvetic:2007ds}. By using Slavnov-Taylor identity one can represent correlators of dressed mean gluons in terms of this integral at two-loop level. 
What kind of integrals will contribute to the scale-independent $Lcc$ correlator at all loops is unclear at present. 
However, using a method of Ref. \cite{Cvetic:2007ds} one can suggest that at higher loop orders in the position space the UD integrals
will survive only. Conformal invariance of the effective action of dressed mean fields in the position space, predicted in  
Refs. \cite{Cvetic:2004kx,Kondrashuk:2004pu,Kang:2004cs,Cvetic:2006kk} corresponds to the property of conformal invariance of 
UD integrals.

In the momentum space it was shown that  UD functions are the only contributions (at least up to three loops) to off-shell four-point correlator 
of gluons that corresponds to four gluon amplitude \cite{Drummond:2006rz,Bern:2005iz}. The conformal invariance of UD functions was used in the momentum space 
to calculate four point amplitude and to classify all possible contributions to it \cite{Bern:2006ew,Nguyen:2007ya}. Later, the conformal symmetry in the momentum space  
appeared on the string side in the Alday-Maldacena approach \cite{Alday:2007hr} in the limit of strong coupling.

The purpose of this paper is to find the position space representation of the ladder diagrams that produce UD functions in the momentum space. 
In this paper we show that Fourier transform of the second UD integral is the second UD integral and that Fourier transform of the first UD function can be related to the second 
UD function. We consider three-point ladder UD integrals and comment four-point ladder UD integrals. In Section 2 we illustrate the idea of the method on an 
example of the simplest diagram. The most important point is that the problem is solved diagrammatically via conformal transformation. Two other solutions to this problem are 
given in Section 3 and Section 4.

\section{First UD triangle diagram}

\begin{figure}[ht]
\begin{minipage}[h]{0.9\linewidth}
\centering\epsfig{file=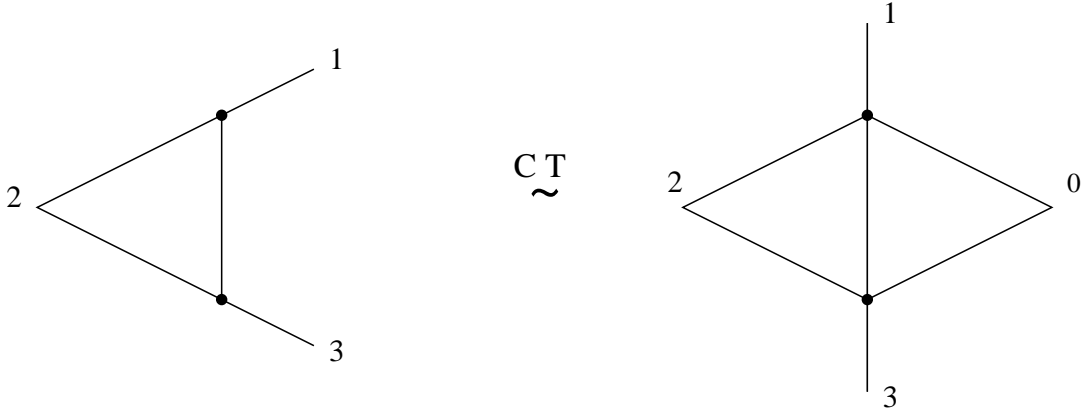,width=\linewidth}
\end{minipage}
\caption{\footnotesize Initial conformal transformation in the position space}
\label{Initial conformal transformation}
\end{figure}

First UD triangle diagram is depicted on the l.h.s. of  Fig. (\ref{Initial conformal transformation}). All the notation used in this paper is the notation of Ref. 
\cite{Cvetic:2006iu}. To calculate it we use conformal transformation. 
The conformal transformation for each vector of the integrand (including the external vectors) is 
\begin{eqnarray}
y_\mu = \frac{y'_\mu}{{y'}^2}, ~~~~ z_\mu = \frac{z'_\mu}{{z'}^2} \label{CT} 
\end{eqnarray}
On the level of equations, the chain of conformal transformations of the l.h.s. of  Fig.(\ref{Initial conformal transformation}) according to Eq. (\ref{CT}) is 
\begin{eqnarray*}
\int~d^4y~d^4z \frac{1}{[2y][yz][2z][3z][1y]} = \\
~[2']^2[3'][1']\int~d^4y'~d^4z'\frac{1}{[2'y'][y'z'][2'z'][3'z'][1'y'][y'][z']} =\\
~[2']^2[3'][1'] \frac{1}{[3'1'][2']^2} \Phi^{(2)}\le \frac{[1'2'][3']}{[3'1'][2']},\frac{[1'][2'3']}{[3'1'][2']}\ri  = \\
\frac{[3'][1']}{[3'1']} \Phi^{(2)}\le \frac{[1'2'][3']}{[3'1'][2']},\frac{[1'][2'3']}{[3'1'][2']}\ri  =  
\frac{1}{[31]} \Phi^{(2)}\le \frac{[12]}{[31]},\frac{[23]}{[31]}\ri .
\end{eqnarray*}
The second row of this chain of transformations looks like the second UD integral in the dual representation of Ref. \cite{Drummond:2006rz}.
It corresponds to the r.h.s. of Fig. (\ref{Initial conformal transformation}). The last line is the conformal transformation back to the initial variables. 
Thus, we have proved the formula
\begin{eqnarray}
\int~d^4y~d^4z \frac{1}{[2y][1y][3z][yz][2z]} = \frac{1}{[31]} \Phi^{(2)}\le \frac{[12]}{[31]},\frac{[23]}{[31]}\ri. \label{first-formula} 
\end{eqnarray}
After making Fourier transformation, we have the following representation for the l.h.s. (definitions of momenta are indicated on Fig. (\ref{one loop})):
\begin{figure}[htb]
\begin{minipage}[h]{0.4\linewidth}
\centering\epsfig{file=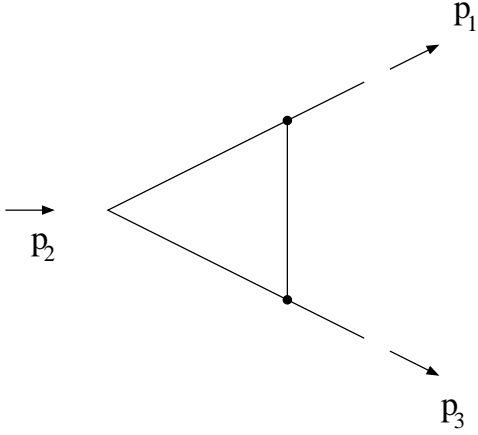,width=\linewidth}
\end{minipage}
\caption{\footnotesize One loop diagram in the momentum space}
\label{one loop}
\end{figure}
\begin{eqnarray*}
\int~d^4y~d^4z \frac{1}{[2y][1y][3z][yz][2z]} = 4\pi^2\int~d^4p_1d^4p_2d^4p_3 d^4r~ \delta(p_1 - p_2 + p_3) \times\\
\times \frac{1}{p_1^2~p_3^2} e^{ip_2x_2} e^{-ip_1x_1} e^{-ip_3x_3} \frac{1}{(r+p_1)^2r^2(r-p_3)^2 } = \\
4\pi^2\int~d^4p_1d^4p_2d^4p_3 ~ \delta(p_1 - p_2 + p_3) e^{ip_2x_2} e^{-ip_1x_1} e^{-ip_3x_3} \times\\
\times \frac{1}{p_2^2~p_1^2~p_3^2} \Phi^{(1)}\le \frac{p_1^2}{p_2^2},\frac{p_3^2}{p_2^2}\ri .
\end{eqnarray*}
Thus, comparing with Eq.(\ref{first-formula}), we can derive the first relation:
\begin{eqnarray}
\frac{1}{[31]} \Phi^{(2)}\le \frac{[12]}{[31]},\frac{[23]}{[31]}\ri = 4\pi^2\int~d^4p_1d^4p_2d^4p_3 ~ \delta(p_1 - p_2 + p_3) \times\no\\
\times e^{ip_2x_2} e^{-ip_1x_1} e^{-ip_3x_3} \frac{1}{p_2^2~p_1^2~p_3^2} \Phi^{(1)}\le \frac{p_1^2}{p_2^2},\frac{p_3^2}{p_2^2}\ri. \label{FF} 
\end{eqnarray}
However, looking at the definition of the UD integrals in Refs. \cite{Usyukina:1992jd,Usyukina:1993ch}, we can write from Eq. (\ref{FF}) another relation:
\begin{eqnarray}
\frac{1}{[31]^2} \Phi^{(2)}\le \frac{[12]}{[31]},\frac{[23]}{[31]}\ri = \frac{1}{16\pi^4}\int~d^4p_1d^4p_2d^4p_3 ~ \delta(p_1 - p_2 + p_3) \times\no\\
\times e^{ip_2x_2} e^{-ip_1x_1} e^{-ip_3x_3} \frac{1}{(p_2^2)^2} \Phi^{(2)}\le \frac{p_1^2}{p_2^2},\frac{p_3^2}{p_2^2}\ri. \label{SF}
\end{eqnarray}
The next two sections demonstrate how to derive the formula (\ref{first-formula}) by other two different methods.

\section{Graphical identity}

\begin{figure}[htb]
\begin{minipage}[b]{0.8\linewidth}
 \centering\epsfig{file=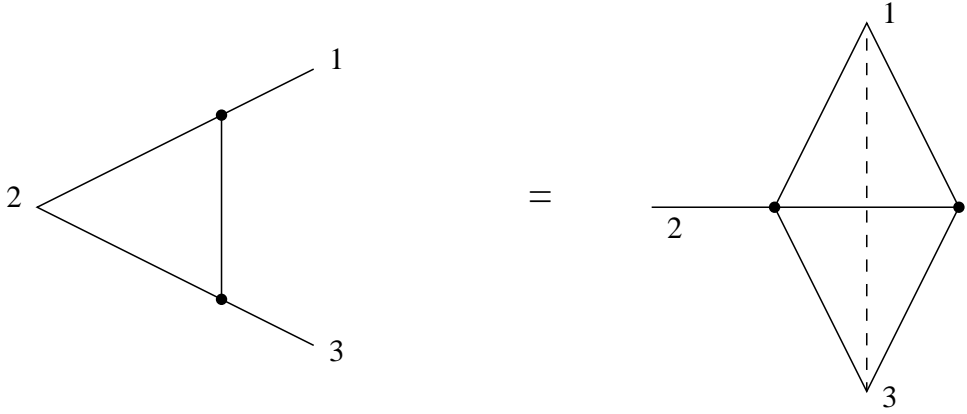,width=\linewidth}
\end{minipage}
\caption{\footnotesize Graphical identity}
\label{Graphical-Id}
\end{figure}

\begin{figure}[ht]
\begin{minipage}[h]{0.9\linewidth}
 \centering\epsfig{file=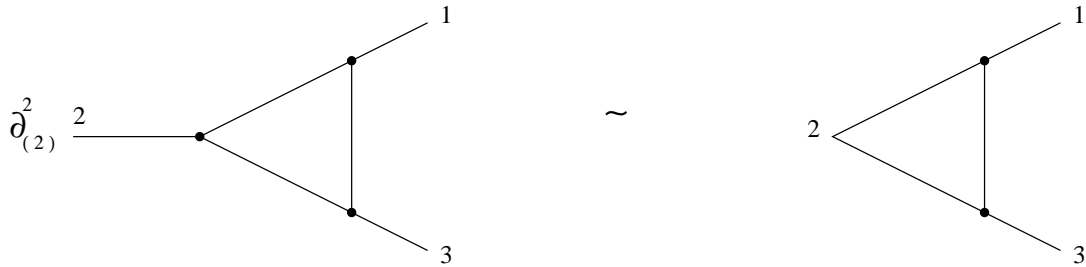,width=\linewidth}
\end{minipage}
\caption{\footnotesize Use of d'Alambertian}
\label{Dalam}
\end{figure}

\begin{figure}[ht]
\begin{minipage}[h]{0.9\linewidth}
 \centering\epsfig{file=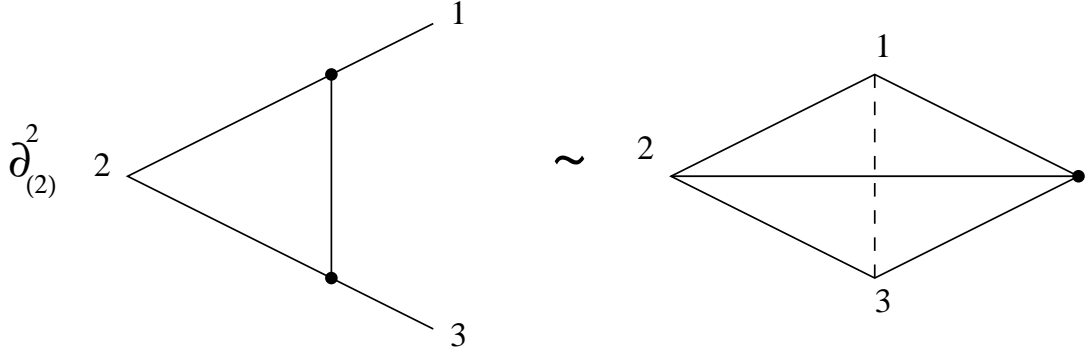,width=\linewidth}
\end{minipage}
\caption{\footnotesize Use of d'Alambertian}
\label{Program}
\end{figure}

\begin{figure}[ht]
\begin{minipage}[h]{1.\linewidth}
 \centering\epsfig{file=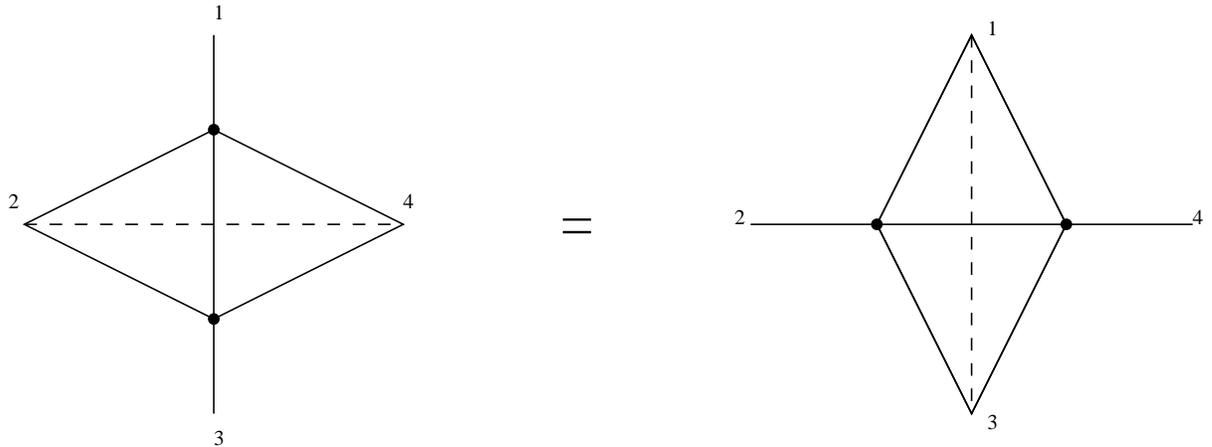,width=\linewidth}
\end{minipage}
\caption{\footnotesize Turning identity. }
\label{Turn-Id}
\end{figure}

\begin{figure}[htb]
\begin{minipage}[b]{0.8\linewidth}
 \centering\epsfig{file=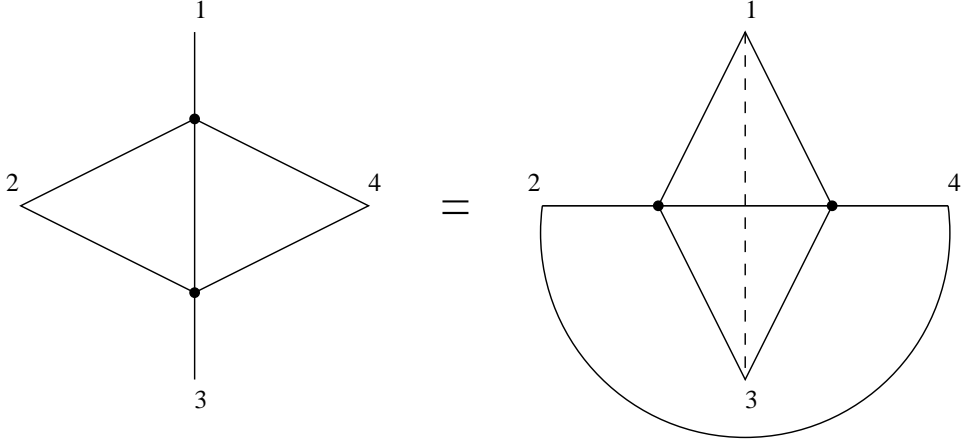,width=\linewidth}
\end{minipage}
\caption{\footnotesize Multiplication by propagator.}
\label{Multiplication}
\end{figure}

\begin{figure}[ht]
\begin{minipage}[h]{0.9\linewidth}
 \centering\epsfig{file=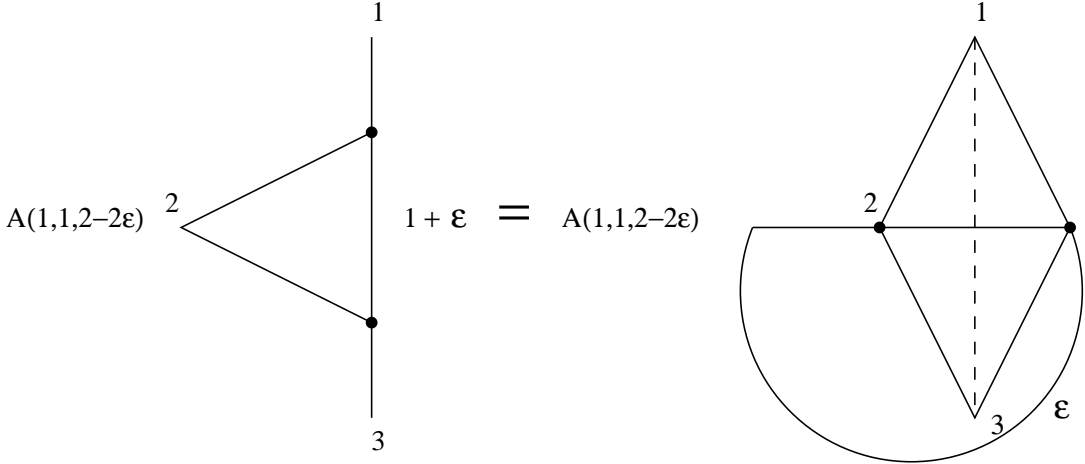,width=\linewidth}
\end{minipage}
\caption{\footnotesize Integration over $x_4.$}
\label{Integration x4}
\end{figure}

First of all, we show validity of the graphical identity of Fig.(\ref{Graphical-Id}). 
This is identity in the position space. We assume integration over internal vertices. This identity can be proved in two ways.  
\\
\noindent {\bf 1.} First way to prove Fig.(\ref{Graphical-Id}) is to use a relation on Fig.(\ref{Dalam}). This is a graphical representation of the equation   (rules of the integration 
are taken from Ref. \cite{Cvetic:2006iu})
\begin{eqnarray*}
\pd^2_{(y)} \frac{1}{[1y]^{1-\e}} = k(\e)\delta^{(4-2\e)}(1y)
\end{eqnarray*}
from Ref \cite{Cvetic:2007ds}. The coefficient between the l.h.s. and the r.h.s. of Fig. (\ref{Dalam}) is $k=-4$ in the number of dimensions $d=4.$  
On the other side, d'Alambertian can travel along the propagator. Computer program of Ref. \cite{Cvetic:2007ds}, written in {\it Mathematica}, produces the equation  
\footnote{ The l.h.s. of (\ref{Identity}) appears on page 24 of Ref.\cite{Cvetic:2007ds}}
\begin{eqnarray}
\pd_{(2)}^2 \int~Dy~Dz \frac{1}{[2y][1y][3z][yz][2z]} = -\frac{4[31]}{[12][23]}~J(1,1,1). \label{Identity}
\end{eqnarray}
This identity is depicted on Fig.(\ref{Program}). The dash lines correspond to the inversed  propagator. The coefficient between the l.h.s. and the r.h.s. of Fig.(\ref{Program}) is $k=-4.$ 
Combining  Fig.(\ref{Dalam}) and Fig.(\ref{Program}) we reproduce the graphical identity of Fig.(\ref{Graphical-Id}). Integration in the internal vertices includes powers 
of $\pi$ due to the definition of Ref. \cite{Cvetic:2006iu}. Since both parts of Fig.(\ref{Graphical-Id}) contain two internal vertices, the identity of Fig.(\ref{Graphical-Id}) 
is valid for a usual four-dimensional measure of integration. 
Thus, we have proved the formula:
\begin{eqnarray}
\int~d^4y~d^4z \frac{1}{[2y][1y][3z][yz][2z]} = [31]\int~d^4y~d^4z \frac{1}{[2y][yz][1y][3y][1z][3z]}.  \label{GI}
\end{eqnarray}
This formula corresponds to the graphical identity of Fig.(\ref{Graphical-Id}).
\\ 
\noindent {\bf 2.} Another way to show validity of Fig.(\ref{Graphical-Id}) is a useful identity of Ref. \cite{Drummond:2006rz} in the position space which can be 
obtained from the property that $\Phi^{(2)}$ function depends on two conformally invariant ratios of spacetime intervals. This representation is valid in 
the position space. The turning identity is re-presented in Fig. (\ref{Turn-Id}).  Historically it appeared  in Ref.\cite{Drummond:2006rz} as a ``dual''  representation 
of the momentum two-loop UD integral  which is not exactly the same as a position representation (the position representation is usual Feynman 
ladder diagram integrated over coordinates of internal vertices). Internal vertices correspond to the momenta that run into the loops.
However, in the dual representation the integrations are done over  ``coordinates'' of the internal vertices too and thus the dual diagram can be considered 
as another Feynman diagram in the position space.   
By multiplying both parts of the turning identity by propagator $[24]$, we have another graphical identity depicted on Fig. (\ref{Multiplication}). 
Integrating both parts over the variable $x_4$ in $d=4-2\e$ dimensions, we obtain the relation re-presented on Fig. \ref{Integration x4},
where 
\begin{eqnarray*}
A(1,1,2-2\e)=\frac{\Gamma^2(1-\e)\Gamma(\e)}{\Gamma(2-2\e)}.
\end{eqnarray*}
Cancelling the coefficient $A(1,1,2-2\e)$ in the both parts and taking the limit $\e \rar 0$, we reproduce Fig.(\ref{Graphical-Id}).

\section{Relation between graphical identity and UD integral}

\begin{figure}[ht]
\begin{minipage}[h]{0.9\linewidth}
\centering\epsfig{file=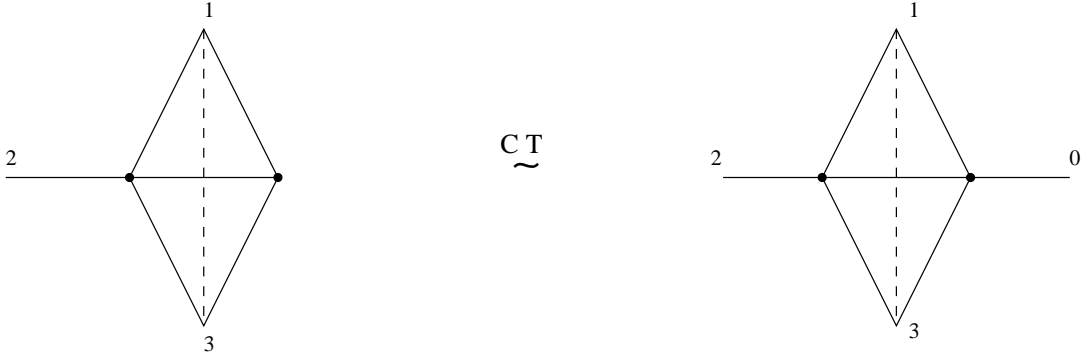,width=\linewidth}
\end{minipage}
\caption{\footnotesize Conformal transformation of the r.h.s. of Fig. \ref{Graphical-Id}}
\label{Conformal transformation}
\end{figure}

\begin{figure}[ht]
\begin{minipage}[h]{0.5\linewidth}
 \centering\epsfig{file=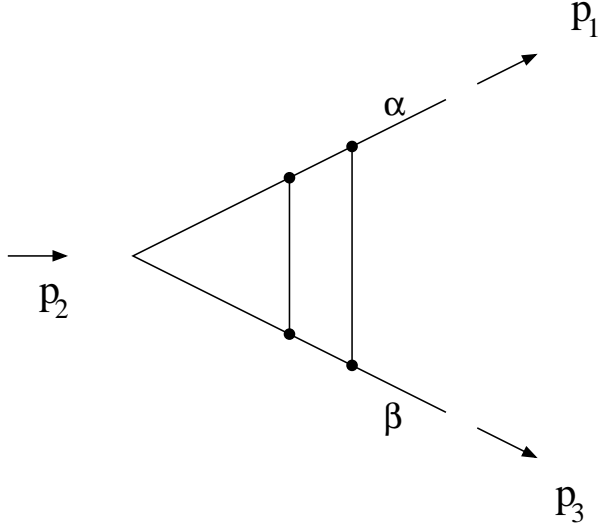,width=\linewidth}
\end{minipage}
\caption{\footnotesize Two-loop diagram}
\label{Two-loop diagram}
\end{figure}

\begin{figure}[htb]
\begin{minipage}[b]{0.9\linewidth}
 \centering\epsfig{file=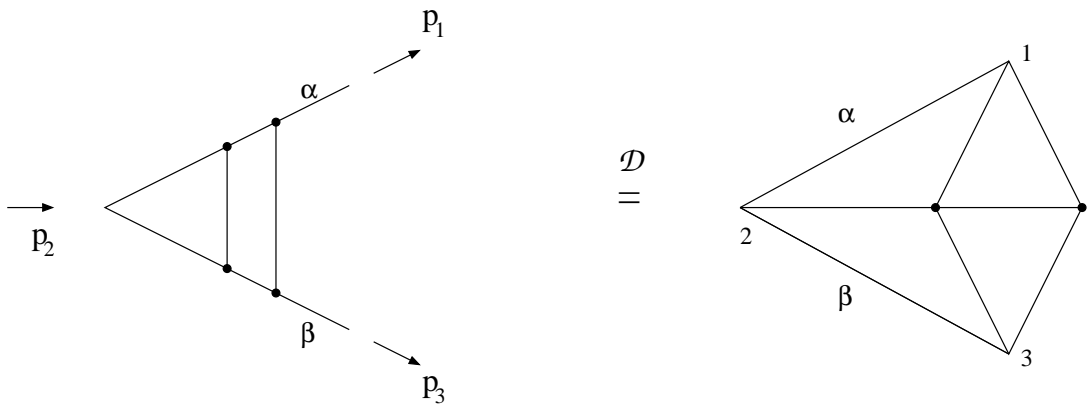,width=\linewidth}
\end{minipage}
\caption{\footnotesize Dual representation}
\label{dual reps}
\end{figure}

\begin{figure}[ht]
\begin{minipage}[h]{0.8\linewidth}
 \centering\epsfig{file=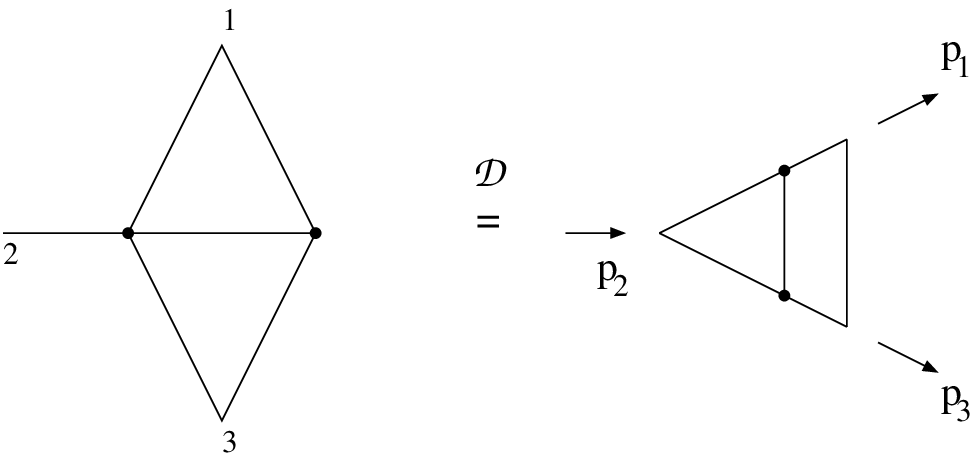,width=\linewidth}
\end{minipage}
\caption{\footnotesize $\alpha =\beta =0$}
\label{alphabeta0}
\end{figure}

After proving Fig.(\ref{Graphical-Id}), we can relate its r.h.s. to UD integrals. This can be done in two ways. \\
\noindent {\bf 1.}   First way is by conformal transformation of  the r.h.s. of Fig. (\ref{Graphical-Id}). Indeed, the integral that corresponds to the r.h.s. of  Fig.(\ref{Graphical-Id}) is 
\begin{eqnarray*}
\int~d^4y~d^4z \frac{1}{[2y][yz][1y][3y][1z][3z]}. 
\end{eqnarray*}
According to the conformal substitution of Eq.(\ref{CT}) the integral of the r.h.s. of Fig. (\ref{Graphical-Id}) can be transformed to 
\begin{eqnarray}
\int~d^4y~d^4z \frac{1}{[2y][yz][1y][3y][1z][3z]} = \no\\
\int~\frac{d^4y'~d^4z'}{[y']^4[z']^4}\frac{[2'][3']^2[1']^2[y']^4[z']^3}{[2'y'][y'z'][1'y'][3'y'][1'z'][3'z']} = \no\\
\int~d^4y'~d^4z'\frac{[2'][3']^2[1']^2}{[2'y'][y'z'][1'y'][3'y'][1'z'][3'z'][z']} \label{rhs}
\end{eqnarray}
The transformation is presented on Fig. (\ref{Conformal transformation}). The r.h.s. of  Fig. (\ref{Conformal transformation}) looks like the second UD integral in the dual 
representation of Ref. \cite{Drummond:2006rz}.
Thus, we can represent Eq. (\ref{rhs}) as 
\begin{eqnarray*}
\int~d^4y'~d^4z'\frac{[2'][3']^2[1']^2}{[2'y'][y'z'][1'y'][3'y'][1'z'][3'z'][z']} = \\
~[2'][3']^2[1']^2 \frac{1}{[3'1']^2[2']} \Phi^{(2)}\le \frac{[1'2'][3']}{[3'1'][2']},\frac{[1'][2'3']}{[3'1'][2']}\ri  = \\
\frac{[3']^2[1']^2 }{[3'1']^2} \Phi^{(2)}\le \frac{[1'2'][3']}{[3'1'][2']},\frac{[1'][2'3']}{[3'1'][2']}\ri  = 
\frac{1}{[31]^2} \Phi^{(2)}\le \frac{[12]}{[31]},\frac{[23]}{[31]}\ri. 
\end{eqnarray*}
The last line is the conformal transformation back to the initial variables of Eq. (\ref{CT}). Thus, we have demonstrated that 
\begin{eqnarray*}
\int~d^4y~d^4z \frac{1}{[2y][yz][1y][3y][1z][3z]} = \frac{1}{[31]^2} \Phi^{(2)}\le \frac{[12]}{[31]},\frac{[23]}{[31]}\ri 
\end{eqnarray*}
Taking into account Eq.(\ref{GI}), we have proved the formula of Eq. (\ref{first-formula}) 
\begin{eqnarray*}
\int~d^4y~d^4z \frac{1}{[2y][1y][3z][yz][2z]} = \frac{1}{[31]} \Phi^{(2)}\le \frac{[12]}{[31]},\frac{[23]}{[31]}\ri. 
\end{eqnarray*}
\noindent {\bf 2.} Second way to demonstrate validity of Eq. (\ref{GI}) does not require the conformal transformation. 
The dual representation for the two-loop diagram on Fig. (\ref{Two-loop diagram}) in the momentum space is given on  Fig. (\ref{dual reps}). 
Thus, for the case $\alpha=\beta=0$ we have the structure of the dual diagram depicted on Fig. (\ref{alphabeta0}). 
However, using the definition of UD functions of Refs. \cite{Usyukina:1992jd,Usyukina:1993ch} and taking into account the relation for dual momenta, one can see that the 
r.h.s. of Fig.(\ref{alphabeta0}) is 
\begin{eqnarray}
\frac{1}{[31]^2}\Phi^{(2)}\le \frac{[12]}{[31]},\frac{[23]}{[31]} \ri. \label{uh}
 \end{eqnarray}
Thus, comparing the l.h.s. of Fig. (\ref{alphabeta0}) with the graphical identity Fig.(\ref{Graphical-Id}) and taking into account 
Eq. (\ref{uh})  we have in the position space for the l.h.s. of  Fig.(\ref{Graphical-Id}) the result 
\begin{eqnarray*}
\frac{1}{[31]}\Phi^{(2)}\le \frac{[12]}{[31]},\frac{[23]}{[31]} \ri. 
\end{eqnarray*}
This coincides with the result in Eq.(\ref{first-formula}).

\section{Conclusion}

We have shown that Fourier transform of the second UD integral is the second UD integral, and that Fourier transforms of the first and the second UD functions are related.
Apart from pure academic interest, this conclusion allows to investigate correlators of ${\cal N} = 4$ supersymmetric Yang-Mills theory in the position space. 
It is useful from the point of view of Slavnov-Taylor identity. This identity will allow to find even maybe yet unknown relations between different types of UD integrals.  
We hope that the property exists above the present consideration, that is for four-point ladder diagrams and at higher loops too. We plan to consider this in future.

\subsection*{Acknowledgments}

We thank Alvaro Vergara for his work with computer graphics for this paper.  
I.K. is grateful to organizers of conference  ``New trends in complex and harmonic analysis'' for giving him wonderful
opportunity to participate in it and for providing financial support of his stay in Bergen. The work of I.K. was supported by  Fondecyt (Chile) project \#1040368, 
and by Departamento de Ciencias B\'asicas de la Universidad  del B\'\i o-B\'\i o, Chill\'an (Chile). A.K. is supported by Fondecyt 
international cooperation project  \#7070064.

\end{document}